\documentclass[fleqn,10pt]{wlscirep}

\usepackage[utf8]{inputenc}
\usepackage[T1]{fontenc}
\usepackage{graphicx}
\usepackage{xcolor}
\usepackage{siunitx}
\usepackage{bbold}
\usepackage{bm}
\usepackage[normalem]{ulem}
\title{Memory-induced alignment of colloidal dumbbells}

\author[1,+]{Karthika Krishna Kumar}
\author[2,+]{Juliana Caspers}
\author[1]{F\'elix Ginot}
\author[2]{Matthias Krüger}
\author[1,*]{Clemens Bechinger}
\affil[1]{Fachbereich Physik, Universit\"{a}t Konstanz, 78457 Konstanz, Germany}
\affil[2]{Institute for Theoretical Physics, Georg-August Universit\"{a}t G\"{o}ttingen, 37073 G\"{o}ttingen, Germany}

\affil[*]{clemens.bechinger@uni-konstanz.de}
\affil[+]{these authors contributed equally to this work}

\begin{abstract}
When a colloidal probe is forced through a viscoelastic fluid which is characterized by a long stress-relaxation time, the fluid is excited out of equilibrium. This is leading to a number of interesting effects including a non-trivial recoil of the probe when the driving force is removed. 
Here, we experimentally and theoretically investigate the transient recoil dynamics of non-spherical particles, i.e., colloidal dumbbells. In addition to a translational recoil of the dumbbells, we also find a pronounced angular reorientation which results from the relaxation of the surrounding fluid. Our findings are in good agreement with a Langevin description based on the symmetries of a director (dumbbell) as well as a microscopic bath-rod model.
Remarkably, we find a frustrated state with amplified fluctuations when the dumbbell is oriented perpendicular to the direction of driving. Our results demonstrate the complex behavior of non-spherical objects within a relaxing environment which are of immediate interest for the motion of externally but also self-driven asymmetric objects in viscoelastic fluids.
\end{abstract}

\begin{document}

\flushbottom
\maketitle
\thispagestyle{empty}

\section{Introduction}

Newtonian liquids typically exhibit relaxation times on the order of microseconds~\cite{dexter1972mechanical} which are significantly shorter than the typical timescales of embedded micron-sized colloidal particles ($\approx$ seconds). 
Accordingly, such liquids remain in equilibrium even when the colloidal particle is subjected to an external driving force. 
The situation is rather different when considering viscoelastic fluids which 
can store and dissipate energy on  timescales on the order of seconds. 
Therefore, pronounced memory effects (non-Markovian behavior) can be expected in such systems due to the slow relaxation of the fluid's mesoscopic microstructure and which has been confirmed using microrheological techniques~\cite{larson_structure_1999,furst_microrheology_2017,chapman2014onset,liu2006microrheology,khan2019optical,weigand2017active,berner_oscillating_2018,jain_two_2021}. 
In particular, so-called recoil experiments,
where first a driving force is applied to a colloidal particle which is then suddenly removed, provide a useful method to explore how the motion of particles is modified when coupled to a slowly relaxing environment~\cite{chapman_nonlinear_2014,gomez-solano_transient_2015,ginot_recoil_2022,caspers_how_2023}. 
Previous studies have revealed rather general double-exponential recoil dynamics which can be quantitatively described by so-called bath particle models, where the response of the fluid is mimicked by harmonically coupled fictitious bath particles~\cite{ginot_recoil_2022,khan2014trajectories}. 
So far, such studies have been only conducted with spherical colloidal particles which produce only axially-symmetric strain fields in the fluid upon shear. 
However, strain fields induced by anisotropic particles are considerably more complex and may not only affect the translational but also the orientational dynamics of such particles within a relaxing viscoelastic fluid.

In this work, we investigate the recoil dynamics of colloidal dumbbells which are driven by an optical tweezers through a viscoelastic fluid. 
In addition to recovering a double-exponential translational recoil, we find a memory-induced alignment (MIA) of the dumbbell along the driving axis. 
This alignment strongly depends on the initial dumbbell orientation relative to the applied driving force and shows a maximum around 45 degrees. 
To explain the origin of this alignment, we derive a nonlinear Langevin description accounting for orientation. 
Based on system symmetries (of a director), this model provides the correct dependence on initial orientation as well as on driving velocity.
Finally, we introduce a microscopic model that fits experimental observations and explains the mechanism behind angle-dependent instabilities. 
Notably, this model requires nonlinear couplings, as the phenomenon of MIA is inherently nonlinear: in linear order, it is forbidden by symmetry.  

\section{Materials and methods}

\begin{figure}
    \centering
   \includegraphics[scale=0.85]{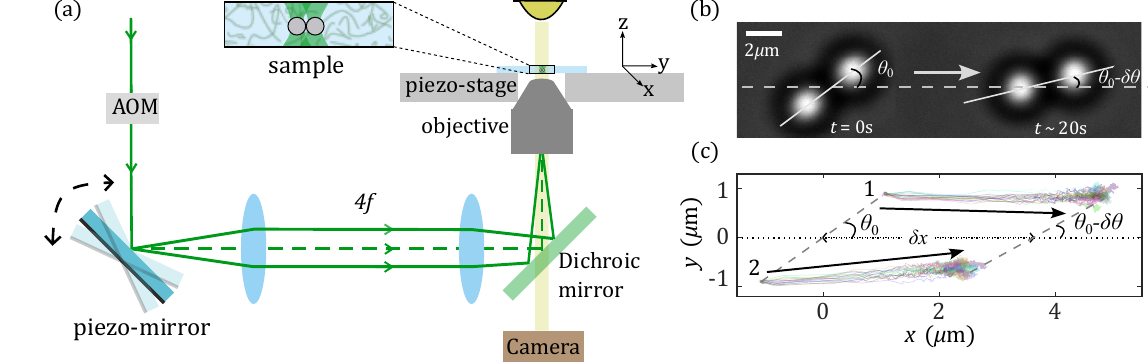}
   \caption{(a) Schematic drawing of the experimental setup.  An oscillating mirror which is mounted onto a piezo device is deflecting a laser beam through an objective into the sample cell where it creates an extended optical trap. (b) Snapshots of a dumbbell at the start $t=0$ and the end of a recoil experiment ($t\sim \SI{20}{\s}$). The initial angle of the dumbbell's long axis $\theta_0$ decreases during the recoil by $\delta\theta$. (c) Corresponding trajectories of the single particles forming the dumbbell particles demonstrate the complex particle motion during recoil. The horizontal dotted line is along the direction of shear. 
   }
    \label{fig1}
\end{figure}

In our study, we use a viscoelastic solution of $\SI{8}{mM}$ equimolar cetyl pyridinium chloride (CPyCl) and sodium salicylate (NaSal). 
A small amount of silica particles with diameter $\SI{2.73}{\mu m}$ is added to the fluid which is contained in a sealed rectangular capillary with $\SI{200}{\mu m}$ height and kept at a constant temperature of $\SI{25}{^\circ C}$. 
At concentrations above the critical micellar concentration ($>\SI{4}{m M}$), the fluid forms giant worm-like micelles also known as 'living polymers' due to dynamic self-healing ~\cite{cates1990statics}.
Opposed to polymer solutions, worm-like micelles within the semi-dilute regime exhibit a particular simple relaxation behavior which is well described by a Maxwell model  with a single relaxation time\cite{spenley1993nonlinear,rehage1988rheological}.
For the formation of dumbbells, we exploit the presence of depletion  forces\cite{ray2013observation} which arise in such systems and which lead to stable aggregates (see SM).
The dumbbells are trapped by extended optical tweezers, generated by a laser beam ($\lambda=\SI{532}{nm}$) focused with a 100x oil immersion objective $(\mathrm{NA}=1.45)$ into the sample cell.
The beam is periodically deflected with a frequency of $\SI{120}{\Hz}$ by a piezo-driven mirror and yields two static (in the lab frame) three-dimensional optical traps with a spacing of almost one particle diameter as shown in Fig.~\ref{fig1}(a). 
The intensity of the beam is controlled by an acousto-optic modulator (AOM) to turn the trap 'on' during shear and 'off' during recoil.
Drag forces on the dumbbells are exerted by a translational piezo-driven stage, which is moving the sample in the $x-$direction.
In general, each experiment is started with the dumbbell inside a stationary trap to guarantee identical starting conditions for each measurement. The dumbbell is then dragged through the fluid with constant velocity $v$ ranging from $0.025-\SI{0.3}{\mu\m/\s}$ and for a time $t_\mathrm{sh}$ of typically $\SI{50}{\s}$ to approach a non-equilibrium steady-state (NESS), where recoil amplitudes weakly depend on $t_\mathrm{sh}$. At time  $t=0$ the trap is turned off, which leads to a recoil motion of the dumbbell, owing to the accumulated strain in the viscoelastic fluid.
We measure the transient trajectories for $\SI{20}{\s}$ until the system fully relaxes.
This protocol is typically repeated over 50 times to yield well-defined averaged recoil curves.
This is necessary due to the stochastic nature of such trajectories, which are sensitive to thermal noise. Using video microscopy, the motion of the dumbbells and individual particles is recorded and the position coordinates are afterward determined using a custom MATLAB algorithm~\cite{crocker1996methods}.  

\begin{figure}
    \centering
  \includegraphics[scale = 1.3]{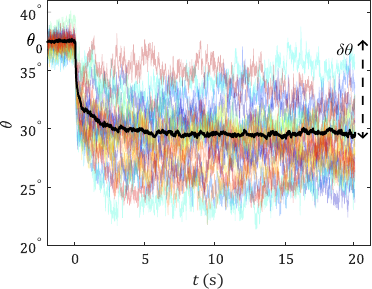}
   \caption{Temporal evolution of angle $\theta$ made by the axis of the dumbbell with recoil direction for different recoil runs (colored lines) and the average curve (thick black line) for $\theta_0\sim40^\circ$, $v = \SI{0.3}{\mu m/s}$ and $t_\mathrm{sh} = \SI{50}{\s}$. The angle made by the dumbbell after release ($t>0$) deviates significantly from the initial angle $\theta_0$ at $t<0$. }
    \label{fig2}
\end{figure}

Typical snapshots of the dumbbell before (trap on) and after a recoil (trap off) and the corresponding trajectories of the particles are shown in Fig.~\ref{fig1}(b\&c) (See also supplementary video 1).
In the following, we will use $\delta x$ for the distance between the initial and final positions of the dumbbell's center of mass (COM), $\theta_0$ for the angle just before release, and $\delta\theta$ for the amplitude of the orientational alignment motion.
As a first observation, the center of mass of the dumbbell exhibits a bi-exponential recoil motion, similar to what was previously reported for simple spherical particles~\cite{ginot_recoil_2022}.
Surprisingly, depending on their position relative to the COM, the particles exhibit different trajectory shapes during the recoil (see Fig.~\ref{fig1}(c)). 
The front particle (1) follows an almost straight trajectory in contrast to the back particle (2) which shows an additional motion in the perpendicular direction, resulting in an oblique trajectory.
This difference results in the angular reorientation of the dumbbell which leads to an alignment of the dumbbell towards the axis of driving.
In Fig.~\ref{fig2} we show the typical evolution of the orientation of the dumbbell during a recoil during several experiments.
The stochastic nature of the system appears clearly with a large spread between individual recoils (faded colored lines).
When averaging over all trajectories we obtain a well-defined average (black line), starting at $\theta_0$ and reaching $\theta_0-\delta\theta$ after $\sim \SI{5}{\s}$. Note, that $\theta_0$ slightly deviates from the orientation of the optical trap due to the presence of drag forces acting on the dumbbell once it is sheared through the fluid (see SM). 
In the following, we will mainly focus on this orientational component of the motion with amplitude  $\delta\theta$. Because it originates (similar to the translational recoil) from the non-Markovian behavior of the fluid, in the remainder of this study we will refer to it as memory-induced alignment (MIA).

\section{Symmetries}\label{chap:symmetries}

\begin{figure}
    \centering
  \includegraphics[scale=0.90]{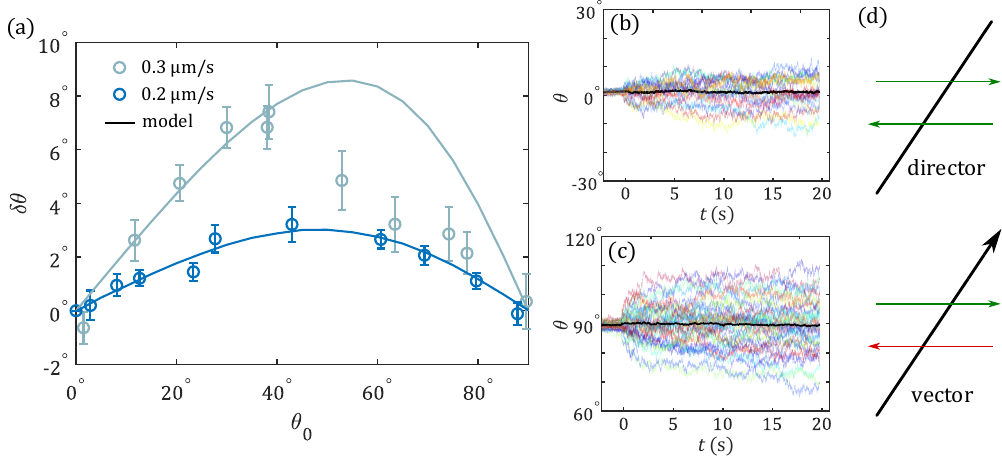}
   \caption{(a) Amplitudes of MIA for fixed shear time $t_\mathrm{sh}=\SI{50}{\s}$ and two different shear velocities, $v=\SI{0.2}{\mu\m/\s}$ (dark blue) and $v=\SI{0.3}{\mu\m/\s}$ (light blue), as a function of initial angle $\theta_0$.  Open symbols correspond to experimental data, the solid line shows simulation results of the model introduced in Fig.~\ref{fig4}(b). The curves show a maximum around $\theta_0=45^\circ$ and decrease to 0 towards $\theta_0=0^\circ$ and $90^\circ$. (b) and (c) show individual recoils (colored lines) and their mean (thick black curve) for $\theta_0=0^\circ$ and $90^\circ$, for fixed shear time $t_\mathrm{sh}=\SI{50}{\s}$ and shear velocity $v=\SI{0.3}{\mu\m/\s}$. Even though the mean curve remains constant in both cases, the individual trajectories show a huge spread at $\theta_0 = 90^\circ$ compared to $\theta_0 = 0^\circ$ signaling an instability when the dumbbell axis is perpendicular to the recoil axis. (d) Sketch of a director (top) and a vector (bottom). While driving left or right is equivalent for a director (green arrows), this is not the case for a vector (green and red arrows lead to different scenarios).  }
    \label{fig3}
\end{figure}

To study the symmetries of the system we vary the initial angle that the dumbbell forms with the recoil direction.
In Fig.~\ref{fig3}(a) we show the resulting MIA amplitudes for a shear velocity $v=\SI{0.2}{\mu\m/\s}$ (dark blue symbols), and $v=\SI{0.3}{\mu\m/\s}$ (light blue symbols) close to the NESS ($t_\mathrm{sh}=\SI{50}{\s}$).
This reveals a non-monotonic behavior with a peak around $45^\circ$, where the alignment between the dumbbell axis and the shear direction is largest.
As expected, larger driving velocities result in larger MIA amplitudes.
For the translational recoil, the initial orientation $\theta_0$ has negligible influence on the amplitudes (see SM).
At $\theta_0 = 0^\circ$ and $\theta_0 = 90^\circ$ the dumbbells are parallel and perpendicular to the recoil direction, respectively,  and we observe no significant change in their orientation.
However, the two complementary orientations induce a very different behavior at the level of individual trajectories.
In Figs.~\ref{fig3}(b) and (c) we show individual recoils for $\theta_0 = 0^\circ$ and $\theta_0=90^\circ$, ($v=\SI{0.3}{\mu\m/\s}$).
One can see that while for $\theta_0 = 0^\circ$ there is almost no change in $\theta$ at the level of individual trajectories (see Fig.~\ref{fig3}(b)), for $\theta_0 = 90^\circ$ there are individual trajectories that deviate significantly from $90^\circ$ in both positive and negative directions (see Fig.~\ref{fig3}(c)).
One hypothesis is that this angle corresponds to a frustrated, meta-stable state, and any thermal fluctuation can lead to MIA in either direction.
The average MIA thus stays null, however, the variance of the trajectories strongly increases.
As will be shown below, the angle of $90^\circ$ indeed corresponds to an unstable configuration in flow. 

Discussing the symmetries of the colloidal dumbbell, we aim for a Langevin description, allowing us to understand the underlying principles of MIA and translational recoil, torque and force.
A dumbbell has an axis of symmetry connecting the centers of the spheres and at the lowest order, it is described by a \textit{director} $\hat{\mathbf{n}}$, i.e., the mentioned axis, and its COM coordinate $\mathbf{x}$. 
The resulting Langevin equations in the phase space spanned by $\hat{\mathbf{n}}$ and $\mathbf{x}$ can be developed based on symmetries. 
We consider the case of a given or prescribed motion, where the stochastic observables are the resulting forces and torques~\cite{Krüger_2017}. 
As motion is prescribed, this description is useful in the presence of a (strong) optical trap. The connection between dynamics with trap on and trap off is non-trivial~\cite{squires_simple_2005,caspers_how_2023,ginot_recoil_2022}, and we do not discuss it here for simplicity.      

Starting with the COM, we note that the force $\mathbf{F}$ acting on the dumbbell must for any $\hat{\mathbf{n}}$ be antisymmetric upon changing the direction of driving, hence scales with odd powers of $\dot{\mathbf x}$.
The lowest order term is therefore linear in $\dot {\mathbf x}$, 
\begin{align}
\begin{split}
      \mathbf{F}(t) = \int_{-\infty}^t \mathrm{d}s\, \boldsymbol{\Gamma}_x(t-s)\cdot \dot{\mathbf{x}}(s) + \boldsymbol{\xi}_F(t)+\dots
      \label{eq:LangevinForce}
      \end{split}
\end{align}
$\boldsymbol{\Gamma}_x$ denotes a memory kernel and $\boldsymbol{\xi}_F$ denotes noise which fulfills FDT, $\langle \boldsymbol{\xi}_F(t)\otimes \boldsymbol{\xi}_F(s)\rangle=k_B T \boldsymbol{\Gamma}_x(|t-s|)$.
The memory kernel has two  distinct contributions due to the mentioned anisotropy of the considered  object,
\begin{align}
    \boldsymbol\Gamma_x(t) = \Gamma_s(t)\mathbb{1} + \Gamma_{n}(t)\hat{\mathbf{n}}\otimes \hat{\mathbf{n}}. 
    \label{eq:GammaF}
\end{align}
The first term is proportional to the identity matrix, resulting in a force parallel to the direction of driving, which resembles the memory function of a spherical object in an isotropic fluid.
The second term is a contribution due to the non-spherical shape of dumbbells, and it is dictated by the symmetry of the director $\hat{\mathbf{n}}$. 
Due to the $\pi/2$ symmetry, the force must scale $\propto \hat{\mathbf{n}}^2$. 
The dots in Eq.~\eqref{eq:LangevinForce} represent higher order terms in driving~\cite{Krüger_2017} in $\dot{\mathbf x}$ and angular velocity $\boldsymbol{\omega} = \hat{\mathbf{n}}\times \dot{\hat{\mathbf{n}}}$.
As the first term in  Eq.~\eqref{eq:LangevinForce} represents the leading term, it is valid if the orientation of the director varies only slowly compared to relaxation times, and $\hat{\mathbf{n}}$ carries no time argument.
We refer to $\hat{\mathbf{n}}$ as initial orientation in our experiments.
Fig.~\ref{fig1}(b) indicate that COM recoil is not perfectly anti-parallel to the direction of trap motion, reflecting the presence of the second term in Eq.~\eqref{eq:GammaF}. As the effect is small, it will not be analyzed in detail.
Allowed (by symmetry) higher order terms include  $\propto \dot{\mathbf{x}}\times \boldsymbol{\omega}$, yielding the recently observed memory induced Magnus force~\cite{cao_memory_2023}. 

Next, we collect contributions to the torque $\boldsymbol{\tau}$ in analogy to Eq.~\eqref{eq:LangevinForce}.
It must be antisymmetric under changes of the direction of angular driving and therefore contains odd powers of $\boldsymbol{\omega}$, i.e., in lowest order the expected linear term,
\begin{align}
   &\boldsymbol{\tau}(t) =  \int_{-\infty}^t \mathrm{d}s \,\boldsymbol\Gamma_\omega(t-s) \cdot\boldsymbol{\omega}(s) 
   +\boldsymbol\xi_\tau(t)+\dots,
    \label{eq:LangevinTheta0}
\end{align}
with $\langle \boldsymbol{\xi}_{\tau}(t)\otimes \boldsymbol{\xi}_{\tau}(s)\rangle=k_B T \boldsymbol{\Gamma}_{\omega}(|t-s|)$, with $\boldsymbol{\Gamma}_{\omega}(|t-s|)\propto\mathbb{1}$ due to symmetries. 
Notably, the set of linear equations \eqref{eq:LangevinForce} and \eqref{eq:LangevinTheta0} do not yield the observed MIA.
In our experiments, $\boldsymbol{\omega}=0$ during drive, 
and no torque builds up according to Eq.~\eqref{eq:LangevinTheta0}.
It is insightful to consider, for a moment, an object with the symmetries of a {\it vector} $\hat{\bf m}$, e.g., dumbbells made of differently sized spheres (compare sketch in Fig.~\ref{fig3}(d)). 
In that case, the leading terms of torque are
   \begin{align}
   &\boldsymbol{\tau}(t) =  \int_{-\infty}^t \mathrm{d}s \,\boldsymbol\Gamma_\omega(t-s) \cdot\boldsymbol{\omega}(s) +
    \int_{-\infty}^t \mathrm{d}s\,\Gamma_m(t-s) (\hat{\bf m} \times \dot{\bf x}(s)) 
   +\boldsymbol\xi_\tau(t)+\dots,
    \label{eq:LangevinThetavec}
\end{align}
and corresponding changes in the noise. 
Eq.~\eqref{eq:LangevinThetavec} shows that the coupling of COM driving and torque already appears in linear order for a vector object. 
In our experiments, the spheres have equal sizes, and that term does not exist. 
The observed MIA thus results from a {\it nonlinear} term. 
It is straightforward to note that, for dumbbells, the direction of MIA must be invariant under an inversion of driving velocity (see Fig.~\ref{fig3}(d)), i.e.~it must be of even orders in $\dot{\mathbf x}$. 
Furthermore, no torque is possible if $\hat{\mathbf{n}}$ and $\dot{\mathbf x}$ are either parallel or perpendicular (compare also  Fig.~\ref{fig3}(a)). 
This yields, to lowest order, a term  $\propto (\dot{\mathbf{x}}\times \hat{\mathbf{n}})(\dot{\mathbf{x}}\cdot \hat{\mathbf{n}})$.
Taking memory into account,
\begin{align}
   &\boldsymbol{\tau}(t) =  \int_{-\infty}^t \mathrm{d}s \,\boldsymbol\Gamma_\omega(t-s) \cdot\boldsymbol{\omega}(s) + \int_{-\infty}^t \mathrm{d}s\int_{-\infty}^t\mathrm{d}s'\, \Gamma^{(2)}_x(t-s,t-s') (\dot{\mathbf{x}}(s) \times \hat{\mathbf{n}})(\dot{\mathbf{x}}(s')\cdot \hat{\mathbf{n}})+\boldsymbol\xi_\tau(t)+\dots
    \label{eq:LangevinTheta}
\end{align}
With Eq.~\eqref{eq:LangevinTheta} being nonlinear, the properties of noise $\boldsymbol{\xi}_\tau(t)$ are nontrivial~\cite{Krüger_2017} and will not be discussed here.

For our specific experimental system, where there is no angular driving and shear velocity is applied in $x$-direction, $\dot{\mathbf{x}}=(\dot{x},0,0)^T$, and using $\hat{\mathbf{n}}=(\cos\theta_0,\sin\theta_0,0)^T$ we find for $\boldsymbol{\tau}=\tau \hat{e}_z$,
\begin{align}
    \tau(t) = \int_{-\infty}^t \mathrm{d}s \int_{-\infty}^t \mathrm{d}s'\, \Gamma^{(2)}_x(t-s,t-s')  \dot{{x}}(s)\dot{{x}}(s')\sin\theta_0\cos\theta_0 +\xi_\theta(t)+\dots .
   \label{eq:LangevinOurSystem}
\end{align}
This functional form $\propto \sin \theta_0 \cos \theta_0$ has a maximum at $45^\circ$ and is symmetric around this angle. 
Indeed, for small driving velocities, our experiments follow well this expected behavior, as seen in Fig.~\ref{fig3}(a) (dark blue).
Note that at higher driving velocities, some deviations are observed for $\theta_0 > 40^\circ$.
The implications for the observed fluctuations and instabilities (Fig.~\ref{fig3}(b) and (c)) will be discussed in detail in Sec.~\ref{chap:fluctuations}.

\section{Microscopic model}\label{chap:model} 

\begin{figure}
    \centering
  \includegraphics[scale = 1.3]{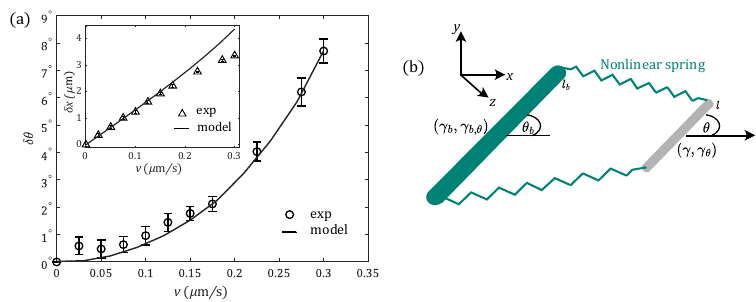}
   \caption{(a) Amplitudes of MIA and translational (inset) recoils as a function of shear velocity $v$ for experiments (open symbols) and simulations (solid line) with fixed initial angle $\theta_0 = 40^\circ$ and shear time $t_\mathrm{sh}=\SI{50}{\s}$. While MIA increases quadratically with $v$, translational recoil is linear in $v$. (b) Sketch of the microscopic model: The colloidal dumbbell is modeled as a rod of length $l$ (gray), coupled to a bath rod (green) of length $l_b$ via a nonlinear spring.}
    \label{fig4}
\end{figure}

Having analyzed the symmetries of MIA and translational recoil, we move on to microscopic modeling~\cite{iubini_aging_2020,baiesi_rise_2021,basu_dynamics_2022,venturelli_memory-induced_2023,demery_non-gaussian_2023}. 
In this intent, we experimentally investigated the dependence on shear velocity $v$ as well as the temporal behavior of recoils. Figure~\ref{fig4}(a) (open symbols) shows the $v$ dependence for initial angle $\theta_0 \sim 40^\circ$, and  $t_\mathrm{sh} = \SI{50}{\s}$ (close to a NESS).
While the MIA amplitude increases quadratically with shear velocity in agreement with our discussion of symmetries in Sec.~\ref{chap:symmetries} (see Eq.~\eqref{eq:LangevinTheta}), the translational component (inset) increases linearly~\cite{ginot_recoil_2022,caspers_how_2023}, for  $v\lesssim\SI{0.2}{\micro\m\per\s}$.
The temporal behavior of recoils are shown in SM. Interestingly, we find that the MIA curves also follow a bi-exponential decay, similar to the translational recoil as found before~\cite{gomez-solano_transient_2015,ginot_recoil_2022,caspers_how_2023}(see SM).
The timescales of translational recoil, $\tau_s^{(t)} = \SI{0.1}{\s}$ and $\tau_l^{(t)} = \SI{2.7}{\s}$, are very similar to the ones observed in \cite{caspers_how_2023}.
This indicates that the timescales are rather independent of the specific size and shape of the probe, but incorporate the relaxation of the fluid.
For the MIA we find slightly shorter timescales of $\tau_s^{(o)}=\SI{0.065}{\s}$ and $\tau_l^{(o)}=\SI{1.29}{\s}$. 

How can MIA be obtained by extending previous bath particle models~\cite{muller_properties_2020,doerries_correlation_2021,ginot_barrier_2022,ginot_recoil_2022,caspers_how_2023}? Naturally, in order to describe dumbbells and angular behavior, the spherical particles of Refs.~\cite{ginot_recoil_2022, caspers_how_2023} acquire now a rod like shape, see Fig.~\ref{fig4}(b). The mentioned two relaxation times require two such bath particles~\cite{ginot_recoil_2022, caspers_how_2023}. However, as the model and analysis for rods is more complicated, we restrict to a single bath rod. While this model naturally only yields single exponential decay of MIA and translational recoils, the challenge and novelty is a microscopic mechanism for the translation-orientation coupling of Eq.~\eqref{eq:LangevinTheta}. More timescales can easily be obtained by adding more bath particles.      

The colloidal dumbbell is thus a rod of length $l$ and COM $\mathbf{x}=(x,y)^T$ and orientation angle $\theta$. $\gamma$ and $\gamma_\theta$ are its translational and orientational friction coefficients, respectively.
For the bath rod, the parameters carry index or superscript $b$. The interaction between the particles differs from previous models~\cite{ginot_recoil_2022} in two ways; i) Springs connect the {\it ends of rods} (see the sketch), $\mathbf{x}_{1,2} = \mathbf{x} \mp l/2 (\cos \theta,\sin \theta)^T$ and $\mathbf{x}_{1,2}^b=\mathbf{x}^b\mp l_b/2 (\cos \theta_b,\sin \theta_b)^T $ respectively, to allow for transmission of torques. ii) While linear for small forces, the springs turn non-linear for larger ones, allowing for the nonlinear coupling  of Eq.~\eqref{eq:LangevinTheta}.  Indeed, using purely harmonic springs does not yield MIA.  
Optical tweezers are modelled by traps of strength $\kappa_x$ and $\kappa_\theta$, respectively, confining COM and angle to $\mathbf{x}_0(t)$ and $\theta_0$.
The final set of stochastic equations describing the model system reads 
(for $i = {1,2}$)
\begin{align}
\begin{split}
    \gamma \dot{\mathbf{x}} &= -\kappa_x(\mathbf{x}-\mathbf{x}_0)-\sum_{i} \nabla_\mathbf{x} V_\mathrm{int}(|\mathbf{x}_i-\mathbf{x}_i^b|) + \boldsymbol\xi \end{split}\\
    \gamma_\theta \dot{\theta} &= - \kappa_\theta (\theta-\theta_0)-\sum_{i}  \partial_{\theta}V_\mathrm{int}(|\mathbf{x}_i-\mathbf{x}_i^b|) + \xi_\theta 
  \\
    \gamma_b \dot{\mathbf{x}}_b &= - \sum_{i}\nabla_{\mathbf{x}_b} V_\mathrm{int}(|\mathbf{x}_i-\mathbf{x}_i^b|) + \boldsymbol\xi_b\\
    \gamma_{b,\theta} \dot{\theta}_b &=  - \sum_{i}\partial_{\theta_b}V_\mathrm{int}(|\mathbf{x}_i-\mathbf{x}_i^b|)+ \xi_{b,\theta},
\end{align}
where $\boldsymbol{\xi}_i$ denotes Gaussian white noise 
\begin{align}
    \langle \boldsymbol{\xi}_i(t)\rangle = 0, \hfill  \langle \boldsymbol{\xi}_i(t)\otimes\boldsymbol{\xi}_j(t')\rangle = 2 k_B T \gamma_i \mathbb{1}\delta_{i,j}\delta(t-t').
\end{align}
The interaction potential is, as mentioned, required to be anharmonic, i.e., 
\begin{align}
    V_\mathrm{int}(x) = \frac{1}{2} \kappa_2 x^2+ \frac{1}{4} \frac{\kappa_4}{l_\mathrm{int}^2} x^4 + \frac{1}{6} \frac{\kappa_6}{l_\mathrm{int}^4} x^6+\dots,\label{eq:Vint}
\end{align}
where the length scale $l_\mathrm{int} = \sqrt{k_B T/\kappa_2}$ ensures that the $\kappa_i$ ($i=2,4,6$) carry the same units of energy divided by length squared. 

The sign of translation-rotation coupling, i.e., the sign of MIA depends on the relative length of $l$ and $l_b$ as well as on the sign of $\kappa_4$. We chose $l_b>l$, which requires, to match experiments, $\kappa_4 <0$, i.e., a weakening at larger extension. To retain stability, a positive $\kappa_6 >0$ is required, and higher order terms, represented by dots in Eq.~\eqref{eq:Vint} are not needed for our purposes.  
For $l=l_b$ there is no MIA.

Solving these equations in simulations, we start by comparing to experimental mean squared displacements (MSD) of freely diffusing dumbbells (absence of trapping), for both COM and angle. This yields the parameters for the linear part of our model. Comparing MIA then fixes $\kappa_4$, noticing that the precise value of $\kappa_6$ is not very relevant. Once fixed (see SM), the model parameters are kept constant throughout the entire paper, i.e., for all different protocols. 

The velocity dependence of MIA amplitude (model results shown as solid lines in Fig.~\ref{fig4}(a)) is very well described by the model, and nicely follows a $\propto v^2$ trend for the velocities shown.
Also for the dependence on $\theta_0$, the model is in very good agreement with experiments and also with expected symmetries, i.e., it follows a $\sin\theta_0 \cos\theta_0$ shape (see Fig.~\ref{fig3}(a), dark blue line). However, for the larger velocity of $v=\SI{0.3}{\mu\m/\s}$, experimental data deviates from that dependence, and this large velocity is clearly beyond the regime of Eq.~\eqref{eq:LangevinTheta} as well as of the model (light blue line). 
This might be expected from Fig.~\ref{fig4}(a) (inset), which shows  the translational recoil amplitude $\delta x$ for varying driving speed $v$ and fixed initial angle $\theta_0=40^\circ$. It clearly shows that the onset of a nonlinear regime for $v >\SI{0.2}{\micro\m\per\s}$  observed in the experiments is not captured by the model.

\begin{figure}[h!]
    \centering
  \includegraphics[scale = 0.65]{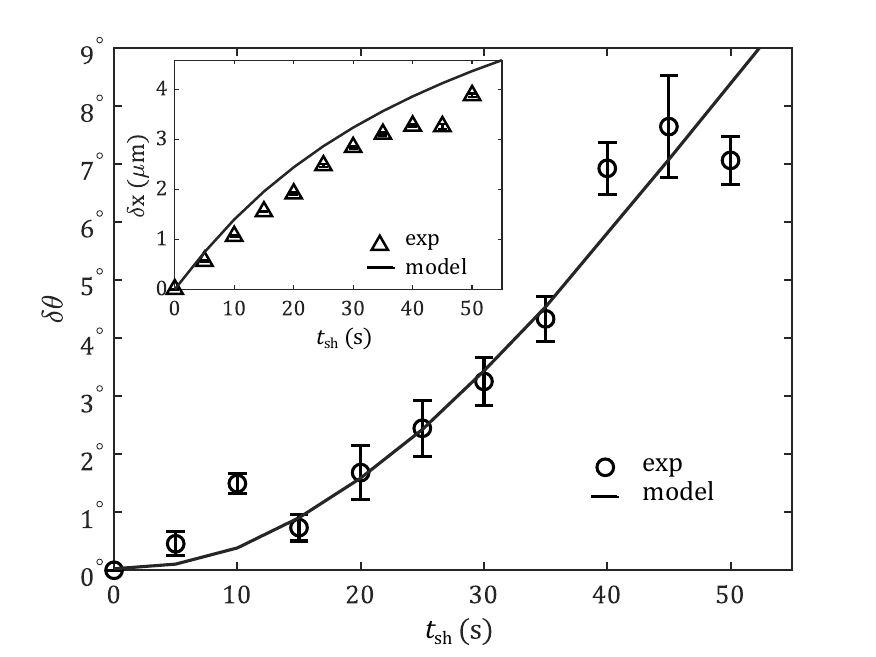}
   \caption{Amplitudes of MIA and translational (inset) recoils for fixed initial angle $\theta_0 = 40^\circ$ and shear velocity $v=\SI{0.3}{\mu\m/\s}$ as a function of shear time $t_\mathrm{sh}$. Open symbols correspond to experimental data and solid lines to simulations. For short times, the MIA amplitude scales $\propto t_\mathrm{sh}^2$, while the translational recoil scales $\propto t_\mathrm{sh}$.}
    \label{fig5}
\end{figure}

To further check the validity of the model, we go beyond the NESS and vary the duration of shear ($t_\mathrm{sh}$) keeping $v=\SI{0.3}{\mu\m/\s}$ and $\theta_0 = 40^\circ$ constant, recalling that the NESS is expected to be reached on timescales that are different from those of recoils~\cite{ginot_recoil_2022}.
The resulting MIA (main graph) and translational (inset) amplitudes from experiments (open symbols) and simulations (solid line) are shown in Fig.~\ref{fig5}.
We observe a quadratic increase in the MIA amplitudes for short values of $t_\mathrm{sh}$, in agreement with simulations, and expected from symmetries.
In contrast, the translational amplitude starts with a linear dependence on $t_\mathrm{sh}$~\cite{ginot_recoil_2022}. Both MIA and translational amplitudes are expected to saturate to a NESS value for large $t_\mathrm{sh}$, which however appears to require longer shearing times. Small deviations between experiments and simulations in the translational recoil amplitude are consistent with previous discrepancies at high shear velocities $v=\SI{0.3}{\mu\m/\s}$ (see Fig.~\ref{fig4}(a)).


\section{Fluctuations and instabilities}\label{chap:fluctuations}

\begin{figure}[h!]
    \centering
  \includegraphics[scale = 0.63]{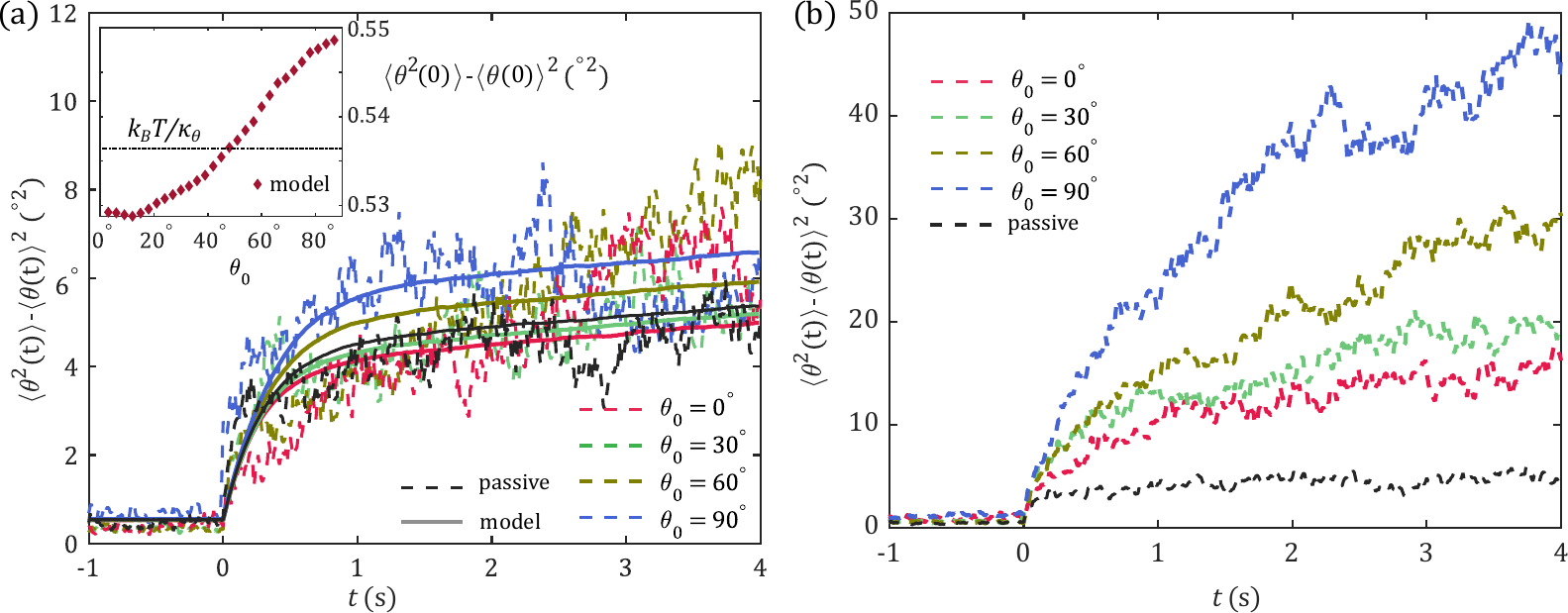}
   \caption{(a) Variance of orientation angle during recoil ($t>0$) for experiments (dashed lines) and simulations (solid lines) for different initial angles (colors) and fixed shear velocity $v=\SI{0.2}{\mu\m/\s}$ and shear time $t_\mathrm{sh}=\SI{50}{\s}$. The black curves correspond to a passive scenario, where dumbbells are released from a stationary trap.
   During recoil, the variance reveals a strong dependence on the initial angle $\theta_0$, with a monotonous increase of amplitude towards $\theta_0 = 90^\circ$.
   Inset: variance of orientation before releasing the trap, obtained from our model, where the passive curve follows the equipartition theorem (black line).  For initial angles $\theta_0 \gtrsim 45^\circ$ fluctuations are amplified while they are suppressed for angles $\theta_0 \lesssim 45^\circ$. (b) Experimental variance during recoil (colored lines) for a shear velocity $v=\SI{0.3}{\mu\m/\s}$. The black line gives the passive scenario.}
    \label{fig6}
\end{figure}

Recoil experiments with colloidal dumbbells have revealed the presence of MIA due to nonlinear coupling with the fluid.  
We analyzed the MIA and translational amplitudes in detail and found quantitative agreement with a microscopic bath-rod model.
However, the error bars in Fig.~\ref{fig3}(a) and the difference in the spread of individual recoil curves for initial angles of $\theta_0 = 0^\circ$ and $\theta_0 = 90^\circ$ in Figs.~\ref{fig3}(b) and (c) for $v=\SI{0.3}{\mu\m/\s}$ already hinted at the importance of fluctuations.
In Fig.~\ref{fig6}(a) we show the variance of orientation angle $\langle \theta^2(t)\rangle-\langle \theta(t)\rangle^2$ for recoil experiments with $v=\SI{0.2}{\mu\m/\s}$, $t_\mathrm{sh}=\SI{50}{\s}$ and different values $\theta_0$ (colored lines).
In addition, we add the passive scenario (black lines), where dumbbells are released from a trap in equilibrium ($v=0$) at $t=0$, for which the variance is naturally independent of $\theta_0$.
From the simulation curves (solid lines), it is clear that the variance increases monotonically with $\theta_0$. 
Notably, for initial orientations of $\theta_0 = 0^\circ$ and $\theta_0=30^\circ$ the variance during recoil is {\it smaller} than the passive curve, while it is larger for $\theta_0 = 60^\circ$ and $\theta_0 = 90^\circ$.
Depending on the angle, a small fluctuation of $\theta$ is amplified or suppressed, so that a larger or smaller variance appears.
This observation can already be understood from the behavior of MIA $\propto \cos \theta_0 \sin\theta_0$ (recall Fig.~\ref{fig3} and  Eq.~\eqref{eq:LangevinOurSystem}). This form has a positive slope for $\theta_0 < 45^\circ$ and thus suppresses fluctuations, and a negative slope for $\theta_0 > 45^\circ$, where fluctuations are amplified. For this small shearing velocity,  experimental curves (dashed lines) in Fig.~\ref{fig6}(a) show a lack of statistics, but, especially for smaller times, an overall increase with initial angle, starting below the equilibrium case, is visible.

To further understand the origin of this instability in our bath-rod model, we consider the variance just before release, i.e., with the trap still on,  as a function of initial angle $\theta_0$ (see inset in Fig.~\ref{fig6}(a)). Here, the equilibrium value is given by $k_BT/\kappa_\theta$ via the equipartition theorem. This value is also met for initial value of $\theta_0\approx 45^\circ$ in the driven case. 
For initial angles above, $\theta_0 \gtrsim 45^\circ$, however, fluctuations are amplified, while they are suppressed for $\theta_0 \lesssim 45^\circ$. The nontrivial shear-induced non-equilibrium fluctuations are therefore already built up during shear and become amplified when the trap is shut off. 

In Fig.~\ref{fig3}(b) and (c) we observed drastic differences in the fluctuations for initial orientations of $\theta_0 = 0^\circ$ and $\theta_0 = 90^\circ$ when a shear velocity of $v=\SI{0.3}{\mu\m/\s}$ has been applied.
Therefore, we also show the variance of orientation angle for recoil experiments for this larger velocity of $v=\SI{0.3}{\mu\m/\s}$, $t_\mathrm{sh} = \SI{50}{\s}$ and different initial angles $\theta_0$ in Fig.~\ref{fig6}(b).
Note that we restrict our analysis to experimental data at this shear velocity, because we already noticed that this large velocity is beyond the range of our theoretical analysis.
In contrast to the recoils at $v=\SI{0.2}{\mu\m/\s}$ the variance for all values of $\theta_0$ is larger than in the passive scenario, and strongly increases with $\theta_0$. In particular, at $t=\SI{4}{\s}$ the variance for initial angles of $\theta_0 = 0^\circ$ and $\theta_0=90^\circ$ differs by a factor of three. These observations provide clear evidence of the presence of pronounced non-equilibrium fluctuations at high shear velocities $v \gtrsim \SI{0.3}{\mu\m/\s}$, when a highly nonlinear regime is entered.

\section{Conclusion}
Using recoil experiments of colloidal dumbbells in a viscoelastic micellar solution, we observed a memory-induced translational and orientational (MIA) recoil.
When varying the shear velocity, we found a quadratic scaling of the MIA amplitude, in agreement with the system's symmetries.
This is a clear signature of the non-linear origin of this phenomenon and contrasts with the linear scaling observed in the translational recoil amplitude.
In parallel, we introduced a phenomenological microscopic bath-rod model, where a torque is created during shear due to the nonlinear coupling between the bath rod and the dumbbell rod. 
This model is in good agreement with our experiments, both, close to the NESS and in the transient regime. Additionally, the MIA shows a strong dependence on the initial orientation of the dumbbell and exhibits a maximal amplitude around $\theta_0 = 45^\circ$.
Using Langevin theory for force and torque, we characterized the dumbbell as a director with $\pi/2$ symmetry.
It then follows that MIA should vanish for parallel ($\theta_0 = 0^\circ$) and perpendicular configurations ($\theta_0 = 90^\circ$), as observed experimentally.
Interestingly, when looking at the MIA amplitude of individual trajectories, these two extreme cases show very different behavior.
The scenario of complete misalignment corresponds to a frustrated state with strongly amplified variance.
More generally, we observed a clear increase in the variance of MIA with the initial angle of driving.
In our microscopic model, this effect is already present during shear and is amplified upon release.
Moreover, when increasing the shear velocity outside of the linear regime, we reported a strong enhancement of the MIA variance.
The amplitude of the measured fluctuations rises well above the thermal diffusion and is clear evidence of the presence of far-from-equilibrium fluctuations.
Further experiments are still required to get more detailed insights into this regime.
Yet, colloidal dumbbells in viscoelastic fluids seem to form a promising system for the study of non-equilibrium fluctuations. 

\section*{Acknowledgements}
This project was funded by the Deutsche Forschungsgemeinschaft (DFG), Grant No. SFB 1432—Project C05. 

\section*{Author contributions}
K.K. performed the experiments under the supervision of F.G. and C.B.; J.C. and M.K. analyzed the symmetries and developed the microscopic model; K.K. and J.C. contributed to writing the manuscript; F.G., M.K. and C.B. provided critical feedback to the analyses and the manuscript; All authors have read and approved the manuscript.

\section*{Competing interests}
The authors declare no competing interests.

\section*{Data availabilty}
The datasets generated during and/or analysed during the current study are available from the corresponding author on reasonable request.


\begin{thebibliography}{10}
\urlstyle{rm}
\expandafter\ifx\csname url\endcsname\relax
  \def\url#1{\texttt{#1}}\fi
\expandafter\ifx\csname urlprefix\endcsname\relax\def\urlprefix{URL }\fi
\expandafter\ifx\csname doiprefix\endcsname\relax\def\doiprefix{DOI: }\fi
\providecommand{\bibinfo}[2]{#2}
\providecommand{\eprint}[2][]{\url{#2}}

\bibitem{dexter1972mechanical}
\bibinfo{author}{Dexter, A.} \& \bibinfo{author}{Matheson, A.}
\newblock \bibinfo{journal}{\bibinfo{title}{The mechanical response of viscous
  liquids}}.
\newblock {\emph{\JournalTitle{Advances in Molecular Relaxation Processes}}}
  \textbf{\bibinfo{volume}{2}}, \bibinfo{pages}{251--318}
  (\bibinfo{year}{1972}).

\bibitem{larson_structure_1999}
\bibinfo{author}{Larson, R.}
\newblock \emph{\bibinfo{title}{The {Structure} and {Rheology} of {Complex}
  {Fluids}}}.
\newblock {EngineeringPro} collection (\bibinfo{publisher}{OUP USA},
  \bibinfo{year}{1999}).

\bibitem{furst_microrheology_2017}
\bibinfo{author}{Furst, E.} \& \bibinfo{author}{Squires, T.}
\newblock \emph{\bibinfo{title}{Microrheology}} (\bibinfo{publisher}{Oxford
  University Press}, \bibinfo{year}{2017}).

\bibitem{chapman2014onset}
\bibinfo{author}{Chapman, C.~D.}, \bibinfo{author}{Lee, K.},
  \bibinfo{author}{Henze, D.}, \bibinfo{author}{Smith, D.~E.} \&
  \bibinfo{author}{Robertson-Anderson, R.~M.}
\newblock \bibinfo{journal}{\bibinfo{title}{Onset of non-continuum effects in
  microrheology of entangled polymer solutions}}.
\newblock {\emph{\JournalTitle{Macromolecules}}} \textbf{\bibinfo{volume}{47}},
  \bibinfo{pages}{1181--1186}, \doiprefix\url{10.1021/ma401615m}
  (\bibinfo{year}{2014}).

\bibitem{liu2006microrheology}
\bibinfo{author}{Liu, J.} \emph{et~al.}
\newblock \bibinfo{journal}{\bibinfo{title}{Microrheology probes length scale
  dependent rheology}}.
\newblock {\emph{\JournalTitle{Phys. Rev. Lett.}}}
  \textbf{\bibinfo{volume}{96}}, \bibinfo{pages}{118104},
  \doiprefix\url{10.1103/PhysRevLett.96.118104} (\bibinfo{year}{2006}).

\bibitem{khan2019optical}
\bibinfo{author}{Khan, M.}, \bibinfo{author}{Regan, K.} \&
  \bibinfo{author}{Robertson-Anderson, R.~M.}
\newblock \bibinfo{journal}{\bibinfo{title}{Optical tweezers microrheology maps
  the dynamics of strain-induced local inhomogeneities in entangled polymers}}.
\newblock {\emph{\JournalTitle{Physical Review Letters}}}
  \textbf{\bibinfo{volume}{123}}, \bibinfo{pages}{038001}
  (\bibinfo{year}{2019}).

\bibitem{weigand2017active}
\bibinfo{author}{Weigand, W.} \emph{et~al.}
\newblock \bibinfo{journal}{\bibinfo{title}{Active microrheology determines
  scale-dependent material properties of chaetopterus mucus}}.
\newblock {\emph{\JournalTitle{PloS one}}} \textbf{\bibinfo{volume}{12}},
  \bibinfo{pages}{e0176732} (\bibinfo{year}{2017}).

\bibitem{berner_oscillating_2018}
\bibinfo{author}{Berner, J.}, \bibinfo{author}{Müller, B.},
  \bibinfo{author}{Gomez-Solano, J.~R.}, \bibinfo{author}{Krüger, M.} \&
  \bibinfo{author}{Bechinger, C.}
\newblock \bibinfo{journal}{\bibinfo{title}{Oscillating modes of driven
  colloids in overdamped systems}}.
\newblock {\emph{\JournalTitle{Nat Commun}}} \textbf{\bibinfo{volume}{9}},
  \bibinfo{pages}{999}, \doiprefix\url{10.1038/s41467-018-03345-2}
  (\bibinfo{year}{2018}).

\bibitem{jain_two_2021}
\bibinfo{author}{Jain, R.}, \bibinfo{author}{Ginot, F.},
  \bibinfo{author}{Berner, J.}, \bibinfo{author}{Bechinger, C.} \&
  \bibinfo{author}{Krüger, M.}
\newblock \bibinfo{journal}{\bibinfo{title}{Two step micro-rheological behavior
  in a viscoelastic fluid}}.
\newblock {\emph{\JournalTitle{J. Chem. Phys.}}}
  \textbf{\bibinfo{volume}{154}}, \bibinfo{pages}{184904},
  \doiprefix\url{10.1063/5.0048320} (\bibinfo{year}{2021}).

\bibitem{chapman_nonlinear_2014}
\bibinfo{author}{Chapman, C.~D.} \& \bibinfo{author}{Robertson-Anderson, R.~M.}
\newblock \bibinfo{journal}{\bibinfo{title}{Nonlinear {Microrheology} {Reveals}
  {Entanglement}-{Driven} {Molecular}-{Level} {Viscoelasticity} of
  {Concentrated} {DNA}}}.
\newblock {\emph{\JournalTitle{Phys. Rev. Lett.}}}
  \textbf{\bibinfo{volume}{113}}, \bibinfo{pages}{098303},
  \doiprefix\url{10.1103/PhysRevLett.113.098303} (\bibinfo{year}{2014}).

\bibitem{gomez-solano_transient_2015}
\bibinfo{author}{Gomez-Solano, J.~R.} \& \bibinfo{author}{Bechinger, C.}
\newblock \bibinfo{journal}{\bibinfo{title}{Transient dynamics of a colloidal
  particle driven through a viscoelastic fluid}}.
\newblock {\emph{\JournalTitle{New J. Phys.}}} \textbf{\bibinfo{volume}{17}},
  \bibinfo{pages}{103032}, \doiprefix\url{10.1088/1367-2630/17/10/103032}
  (\bibinfo{year}{2015}).

\bibitem{ginot_recoil_2022}
\bibinfo{author}{Ginot, F.} \emph{et~al.}
\newblock \bibinfo{journal}{\bibinfo{title}{Recoil experiments determine the
  eigenmodes of viscoelastic fluids}}.
\newblock {\emph{\JournalTitle{New J. Phys.}}} \textbf{\bibinfo{volume}{24}},
  \bibinfo{pages}{123013}, \doiprefix\url{10.1088/1367-2630/aca8c7}
  (\bibinfo{year}{2022}).

\bibitem{caspers_how_2023}
\bibinfo{author}{Caspers, J.} \emph{et~al.}
\newblock \bibinfo{journal}{\bibinfo{title}{How are mobility and friction
  related in viscoelastic fluids?}}
\newblock {\emph{\JournalTitle{J. Chem. Phys.}}}
  \textbf{\bibinfo{volume}{158}}, \bibinfo{pages}{024901},
  \doiprefix\url{10.1063/5.0129639} (\bibinfo{year}{2023}).

\bibitem{khan2014trajectories}
\bibinfo{author}{Khan, M.} \& \bibinfo{author}{Mason, T.~G.}
\newblock \bibinfo{journal}{\bibinfo{title}{Trajectories of probe spheres in
  generalized linear viscoelastic complex fluids}}.
\newblock {\emph{\JournalTitle{Soft matter}}} \textbf{\bibinfo{volume}{10}},
  \bibinfo{pages}{9073--9081} (\bibinfo{year}{2014}).

\bibitem{cates1990statics}
\bibinfo{author}{Cates, M.~E.} \& \bibinfo{author}{Candau, S.~J.}
\newblock \bibinfo{journal}{\bibinfo{title}{Statics and dynamics of worm-like
  surfactant micelles}}.
\newblock {\emph{\JournalTitle{J. Phys. Condens. Matter}}}
  \textbf{\bibinfo{volume}{2}}, \bibinfo{pages}{6869} (\bibinfo{year}{1990}).

\bibitem{spenley1993nonlinear}
\bibinfo{author}{Spenley, N.}, \bibinfo{author}{Cates, M.} \&
  \bibinfo{author}{McLeish, T.}
\newblock \bibinfo{journal}{\bibinfo{title}{Nonlinear rheology of wormlike
  micelles}}.
\newblock {\emph{\JournalTitle{Phys. Rev. Lett.}}}
  \textbf{\bibinfo{volume}{71}}, \bibinfo{pages}{939} (\bibinfo{year}{1993}).

\bibitem{rehage1988rheological}
\bibinfo{author}{Rehage, H.} \& \bibinfo{author}{Hoffmann, H.}
\newblock \bibinfo{journal}{\bibinfo{title}{Rheological properties of
  viscoelastic surfactant systems}}.
\newblock {\emph{\JournalTitle{The Journal of Physical Chemistry}}}
  \textbf{\bibinfo{volume}{92}}, \bibinfo{pages}{4712--4719}
  (\bibinfo{year}{1988}).

\bibitem{ray2013observation}
\bibinfo{author}{Ray, D.} \& \bibinfo{author}{Aswal, V.}
\newblock \bibinfo{journal}{\bibinfo{title}{Observation of adsorption versus
  depletion interaction for charged silica nanoparticles in the presence of
  non-ionic surfactant}}.
\newblock {\emph{\JournalTitle{Journal of Physics: Condensed Matter}}}
  \textbf{\bibinfo{volume}{26}}, \bibinfo{pages}{035102}
  (\bibinfo{year}{2013}).

\bibitem{crocker1996methods}
\bibinfo{author}{Crocker, J.~C.} \& \bibinfo{author}{Grier, D.~G.}
\newblock \bibinfo{journal}{\bibinfo{title}{Methods of digital video microscopy
  for colloidal studies}}.
\newblock {\emph{\JournalTitle{J. Colloid Interface Sci.}}}
  \textbf{\bibinfo{volume}{179}}, \bibinfo{pages}{298--310}
  (\bibinfo{year}{1996}).

\bibitem{Krüger_2017}
\bibinfo{author}{Krüger, M.} \& \bibinfo{author}{Maes, C.}
\newblock \bibinfo{journal}{\bibinfo{title}{The modified {Langevin} description
  for probes in a nonlinear medium}}.
\newblock {\emph{\JournalTitle{J. Phys.: Condens. Matter}}}
  \textbf{\bibinfo{volume}{29}}, \bibinfo{pages}{064004},
  \doiprefix\url{10.1088/1361-648X/29/6/064004} (\bibinfo{year}{2016}).

\bibitem{squires_simple_2005}
\bibinfo{author}{Squires, T.~M.} \& \bibinfo{author}{Brady, J.~F.}
\newblock \bibinfo{journal}{\bibinfo{title}{A simple paradigm for active and
  nonlinear microrheology}}.
\newblock {\emph{\JournalTitle{Physics of Fluids}}}
  \textbf{\bibinfo{volume}{17}}, \bibinfo{pages}{073101},
  \doiprefix\url{10.1063/1.1960607} (\bibinfo{year}{2005}).

\bibitem{cao_memory_2023}
\bibinfo{author}{Cao, X.} \emph{et~al.}
\newblock \bibinfo{title}{Memory induced {Magnus} effect}
  (\bibinfo{year}{2023}).
\newblock \bibinfo{note}{ArXiv:2303.07416 [cond-mat]}.

\bibitem{iubini_aging_2020}
\bibinfo{author}{Iubini, S.}, \bibinfo{author}{Baiesi, M.} \&
  \bibinfo{author}{Orlandini, E.}
\newblock \bibinfo{journal}{\bibinfo{title}{Aging of living polymer networks: a
  model with patchy particles}}.
\newblock {\emph{\JournalTitle{Soft Matter}}} \textbf{\bibinfo{volume}{16}},
  \bibinfo{pages}{9543--9552}, \doiprefix\url{10.1039/D0SM01391A}
  (\bibinfo{year}{2020}).

\bibitem{baiesi_rise_2021}
\bibinfo{author}{Baiesi, M.}, \bibinfo{author}{Iubini, S.} \&
  \bibinfo{author}{Orlandini, E.}
\newblock \bibinfo{journal}{\bibinfo{title}{The rise and fall of branching: {A}
  slowing down mechanism in relaxing wormlike micellar networks}}.
\newblock {\emph{\JournalTitle{The Journal of Chemical Physics}}}
  \textbf{\bibinfo{volume}{155}}, \bibinfo{pages}{214905},
  \doiprefix\url{10.1063/5.0072374} (\bibinfo{year}{2021}).

\bibitem{basu_dynamics_2022}
\bibinfo{author}{Basu, U.}, \bibinfo{author}{Démery, V.} \&
  \bibinfo{author}{Gambassi, A.}
\newblock \bibinfo{journal}{\bibinfo{title}{Dynamics of a colloidal particle
  coupled to a {Gaussian} field: from a confinement-dependent to a non-linear
  memory}}.
\newblock {\emph{\JournalTitle{SciPost Physics}}}
  \textbf{\bibinfo{volume}{13}}, \bibinfo{pages}{078},
  \doiprefix\url{10.21468/SciPostPhys.13.4.078} (\bibinfo{year}{2022}).

\bibitem{venturelli_memory-induced_2023}
\bibinfo{author}{Venturelli, D.} \& \bibinfo{author}{Gambassi, A.}
\newblock \bibinfo{title}{Memory-induced oscillations of a driven particle in a
  dissipative correlated medium} (\bibinfo{year}{2023}).
\newblock \bibinfo{note}{ArXiv:2304.09684 [cond-mat]}.

\bibitem{demery_non-gaussian_2023}
\bibinfo{author}{Démery, V.} \& \bibinfo{author}{Gambassi, A.}
\newblock \bibinfo{title}{Non-{Gaussian} fluctuations of a probe coupled to a
  {Gaussian} field}, \doiprefix\url{10.48550/arXiv.2307.07721}
  (\bibinfo{year}{2023}).
\newblock \bibinfo{note}{ArXiv:2307.07721 [cond-mat]}.

\bibitem{muller_properties_2020}
\bibinfo{author}{Müller, B.}, \bibinfo{author}{Berner, J.},
  \bibinfo{author}{Bechinger, C.} \& \bibinfo{author}{Krüger, M.}
\newblock \bibinfo{journal}{\bibinfo{title}{Properties of a nonlinear bath:
  experiments, theory, and a stochastic {Prandtl}–{Tomlinson} model}}.
\newblock {\emph{\JournalTitle{New J. Phys.}}} \textbf{\bibinfo{volume}{22}},
  \bibinfo{pages}{023014}, \doiprefix\url{10.1088/1367-2630/ab6a39}
  (\bibinfo{year}{2020}).

\bibitem{doerries_correlation_2021}
\bibinfo{author}{Doerries, T.~J.}, \bibinfo{author}{Loos, S. A.~M.} \&
  \bibinfo{author}{Klapp, S. H.~L.}
\newblock \bibinfo{journal}{\bibinfo{title}{Correlation functions of
  non-{Markovian} systems out of equilibrium: analytical expressions beyond
  single-exponential memory}}.
\newblock {\emph{\JournalTitle{J. Stat. Mech.}}}
  \textbf{\bibinfo{volume}{2021}}, \bibinfo{pages}{033202},
  \doiprefix\url{10.1088/1742-5468/abdead} (\bibinfo{year}{2021}).

\bibitem{ginot_barrier_2022}
\bibinfo{author}{Ginot, F.}, \bibinfo{author}{Caspers, J.},
  \bibinfo{author}{Krüger, M.} \& \bibinfo{author}{Bechinger, C.}
\newblock \bibinfo{journal}{\bibinfo{title}{Barrier {Crossing} in a
  {Viscoelastic} {Bath}}}.
\newblock {\emph{\JournalTitle{Phys. Rev. Lett.}}}
  \textbf{\bibinfo{volume}{128}}, \bibinfo{pages}{028001},
  \doiprefix\url{10.1103/PhysRevLett.128.028001} (\bibinfo{year}{2022}).

\end{thebibliography}

\end{document}